\documentclass[preprint,12pt]{elsarticle}

\usepackage{amssymb}

\usepackage{caption}
\usepackage{color}
\usepackage{url}

\newcounter{bla}

\journal{Computer Physics Communications}

\begin{document}

\begin{frontmatter}



\title{UKRmol-scripts: a Perl-based system for the automated operation of the photoionization and electron/positron scattering suite UKRmol+}


\author[ITP]{Karel Houfek}
\author[ITP]{Jakub Benda}
\author[ITP]{Zdeněk Mašín}
\author[ITP,MBI]{Alex Harvey}
\author[ITP]{Thomas Meltzer}
\author[OU]{Vincent Graves}
\author[OU]{Jimena D. Gorfinkiel}

\address[ITP]{Institute of Theoretical Physics, Faculty of Mathematics and Physics, Charles University, V Holešovičkách 2, 180 00 Praha 8, Czech Republic}
\address[OU]{School of Physical Sciences, The Open University, Walton Hall, MK76AA, Milton Keynes, United Kingdom}
\address[MBI]{Max Born Institute, Max-Born-Straße 2A, 12489 Berlin}

\begin{abstract}
UKRmol-scripts is a set of Perl scripts to automatically run the UKRmol+ codes, a complex software suite based on the R-matrix method to calculate fixed-nuclei photoionization and electron- and positron-scattering  for polyatomic molecules. Starting with several basic parameters, the scripts operatively produce all necessary input files and run all codes for electronic structure and scattering calculations as well as gather the more frequently required outputs. The scripts provide a simple way to run such calculations for many molecular geometries concurrently and collect the resulting  data for easier post-processing and visualization. We describe the structure of the scripts and the input parameters as well as provide examples for photoionization and electron and positron collisions with molecules. The codes are freely available from Zenodo.

%
%
%

\end{abstract}


\end{frontmatter}



{\bf PROGRAM SUMMARY}

\begin{small}
\noindent
{\em Program Title:} UKRmol-scripts \\
{\em CPC Library link to program files:} (to be added by Technical Editor) \\
{\em Developer's repository link:} \url{https://gitlab.com/Uk-amor/UKRMol/UKRmol-scripts} \\
{\em Code Ocean capsule:} (to be added by Technical Editor)\\
{\em Licensing provisions:} GPLv3  \\
{\em Programming language:} Perl \\
{\em Supplementary material:}  None                               \\
{\em Nature of problem:} Performing \textit{ab initio}  photoionization and low energy electron- and positron-scattering on polyatomic molecules requires selecting and setting a significant number of input parameters in order to model the physics in a numerically accurate way. These scripts streamline setting up and analyzing calculations based on the R-matrix method \cite{Burkebook} using the UKRmol+ suite \cite{UKRmolp}.
\\
{\em Solution method:} The scripts provide automatic generation of input files and execution of programs from the UKRmol+ suite \cite{UKRmolp} using a number of parameters that describe both physical models and machine-dependent settings, and also outputs of the previous programs in the suite which are automatically read and analyzed. The resulting output files are then post-processed to collect target and scattering data for further analysis and simple plotting.\\
{\em Additional comments:} The scripts should be used with releases 3.2 of UKRmol-in and UKRmol-out \cite{zenodo} although they are compatible with earlier versions.\\
   \\

\end{small}

\section{ Introduction}

The increasing  computational power available to Atomic,  Molecular, and Optical (AMO) physicists has made it possible to run ever more accurate  calculations to investigate photon and lepton-induced processes. This has led to more sophisticated, and therefore complex, software suites that require the user to provide input for several interrelated programs which need to be run, serially or in parallel, in a specific order. This is particularly true when the systems under study are molecules and computational chemistry approaches are used to generate the electronic wave functions.

One of the most established AMO codes for ab initio description of the electronic continuum of molecules is the UKRmol+ suite \cite{UKRMOL+}. It is an  implementation of the time-independent  R-matrix method \cite{B2011,tennyson} to treat molecular photoionization, low-energy electron and positrons scattering for molecules and to provide input to the R-matrix with time-dependence codes RMT \cite{RMTMOL}.  RMT models atoms and molecules in intense, ultrashort, arbitrarily-polarized laser pulses \cite{RMT,RMT_arb}. Among the processes that have been modeled for atoms using RMT are: attosecond transient absorption spectroscopy \cite{SFI_ATAS}, reconstruction of attosecond beating by interference of two-photon transitions (RABBITT) with skewed laser polarisation \cite{partial_wave_meter}, and XUV-initiated high harmonic generation  (HHG) \cite{brown_xuvhhg}. 

The UKRmol+ and RMT suites have so far been used  to  study one-photon and strong-field ionization and  RABBIT time-delays for some diatomic and triatomic molecules \cite{PRA-RMT-molec,PRA_RABBIT_TI}. In addition,  UKRmol+ implements a recently developed time-independent approach to calculate transition moments for the multiphoton ionization, for any number of photons,  of molecules in the perturbative regime \cite{multiphoton}. The approach can be applied to the calculation of RABBIT time-delays \cite{PRA_RABBIT_TI}, photoionization cross sections, and laboratory-frame photoelectron angular distributions.

Running the UKRmol+ suite for different types of physical processes requires, in principle, producing input files for between 5 and 10+ programs, defining both the models that will be used to describe the physics and the technical characteristics of the particular run, from  executables to use, to the number of nodes, tasks, etc. The generation of input files for several consequently executed programs, in particular, is prone to human error: inconsistencies can lead to calculations that either fail or, worse still, produce physically meaningless results and in the process  unnecessarily use up computing resources. Moreover, when calculations for many molecular geometries are needed, the manual generation of all inputs can also be tedious and time-consuming.

These problems were recognized and addressed for the UKRmol+ suite and its  predecessor (the UKRmol suite \cite{epjd_ukrmol}) around 10 years ago. A set of Perl scripts, named UKRmol-scripts, was developed to simplify the running of the suite and make it more accessible. The scripts produce the input for all the programs in the suite, run the codes in the appropriate order and collect the data in a user-friendly format for analyzing, plotting, and comparing. They are significantly more straightforward to master  by  new users than  the UKRmol+ suite itself but allow sufficient sophistication to enable cutting-edge calculations to be performed. The scripts  are reasonably straightforward  to modify, enabling users to  include additional input if required, and to update as new functionality is introduced in the UKRmol+ suite. The scripts are available for download from Zenodo \cite{ukrmol-scripts-zenodo}.

There have been other approaches to make UKRmol+, and other AMO codes, user-friendly. For example, a commercial package, QEC \cite{QEC}, is available to drive UKRmol+. A wider initiative, to make a range of AMO codes available to users, is the AMOSGateway \cite{amos_gateway,AMPGateway-art}: this is a free, easy-to-use platform that gives access to a number of software suites that calculate cross sections for photoionization, electron and positron  scattering,  and solutions of the time-dependent Schr\"odinger equation in the presence of strong and ultrafast electromagnetic fields.

In this paper, we describe  the UKRmol-scripts in detail. Our aim is to encourage more researchers to  use them together with the UKRmol+ suite and exploit the software capabilities to treat photoionization  processes, in particular those induced by ultrashort laser pulses. Section~\ref{sec:ukrmol+} briefly describes the UKRmol+ suite; Section~\ref{sec:scripts} details the use of the UKRmol-scripts, Sections~\ref{sec:example_scat}, \ref{sec:photo} and \ref{sec:rmt} describe the specifics of running scattering, photoionization, and RMT input calculations respectively, including a description of the examples provided with the release. Section~\ref{sec:selectedmodels} describes some newer models to generate the target and all system configurations. Finally in Section~\ref{sec:conclusions}, we present some brief conclusions.

\section{The UKRmol+ suite}\label{sec:ukrmol+}

A detailed description of the UKRmol+ suite can be found in \cite{UKRMOL+}. The codes are freely available for download from Zenodo \cite{ukrmol-in-zenodo, ukrmol-out-zenodo}. Releases include a set of CMake scripts for the configuration, compilation, testing, and installation of the suite. The suite makes use of the GBTOlib library \cite{gbtolib} for computing the required integrals.

The basis of the R-matrix approach is the separation of space into two regions. The boundary between them is a sphere of radius $a$. In the outer region, correlation as well as exchange  between the  unbound particle (when it is an electron) and the molecular electrons can be neglected. In the inner region, these effects need to be taken into account. Quantum chemistry approaches, including the use of Gaussian-type orbital (GTO) basis sets and configuration interaction, are employed to describe both the neutral or cationic bound molecular target states and those of the (anionic or neutral) target + unbound particle system. In the outer region, the interaction potential between the target and the unbound particle is modelled in terms of a multipole expansion.

The suite requires external quantum chemistry codes, such as Molpro \cite{MOLPRO_brief}, Molcas \cite{Molcas} or Psi4 \cite{PSI4}, to generate bound molecular orbitals. The unbound particle is described in the inner region through the use of  purpose-built  Gaussian and/or B-spline basis sets, GTOs and BTOs respectively, to represent the continuum \cite{UKRMOL+}.  The configuration state functions used to describe both the target and the whole system need to be carefully selected to accurately describe the physics of the problem while ensuring the calculation remains tractable.

The suite allows the user to determine  quantities associated with photoionization  (cross sections, asymmetry parameters, Dyson orbitals, etc.) including the transition dipole moments for the R-matrix states. The latter, together with some additional information, can be packaged as input data  to run  RMT calculations. Alternatively, the user can compute electronically elastic and inelastic integral cross sections for electron and positron scattering, as well as resonance energies and lifetimes for the former collision. Output from the suite, in the form of scattering data (K, S, or T-matrices) can be used, for example,  to determine angular differential cross sections.   

We strongly recommend reading this paper in conjunction with the paper on the UKRmol+ suite \cite{UKRMOL+} where all programs mentioned below are described in detail. 

\section{The UKRmol-scripts}\label{sec:scripts}

The UKRmol+ scripts consist of three Perl libraries (modules) placed in the directory \texttt{lib} and several Perl script files in the directory \texttt{scripts}. The module \texttt{dirfile.pm} contains general routines for dealing with files and directories, \texttt{ukrmollib.pm} specific routines for working with the UKRmol+ suite, and \texttt{MultiSpace.pm} routines for multiple active space (MAS) support. The driving script \texttt{main.pl}, which operatively prepares inputs, runs programs and reads and collects their outputs, is called from the command line as
\begin{center}
\texttt{perl main.pl dirs.pl config.pl geometry.pl model.pl}
\end{center}
The auxiliary scripts \texttt{dirs.pl}, ..., \texttt{model.pl} contain all necessary settings, options, and parameters in the form of Perl hash arrays for a specific calculation and are described in detail below.
Moreover, the scripts use the Perl library \texttt{ForkManager.pm} \cite{ForkManager} for a simple parallelization (over target geometries and/or symmetries) if requested.

To run  the UKRmol+ scripts successfully, it is first necessary to modify the settings in \texttt{dirs.pl}. Here, paths and directories are specified where the executable files of the UKRmol+ suite, Molpro, Molcas, or Psi4 are installed and where a copy of the directories \texttt{input.templates} and \texttt{basis.sets}, which are a part of the UKRmol-scripts package, are placed. The \texttt{input.templates} directory provides templates used by the scripts to prepare the inputs for UKRmol+ programs where all strings of the form \texttt{>>>...<<<} are replaced with appropriate values determined from the parameters and available outputs. The \texttt{basis.sets} directory contains standard GTO basis files for both bound molecular and continuum orbitals. The latter are generated separately using other programs from the UKRmol+ suite which are not driven by the scripts. The path to the directory \texttt{lib} can also  be specified either by setting the environment variable \texttt{PERL5LIB} or using -I when running the Perl command. 

The other three scripts \texttt{config.pl}, \texttt{geometry.pl}, and \texttt{model.pl} are used to set up a specific calculation. In \texttt{config.pl}, the user can specify options regarding the overall run of the scripts (whether to run a scattering or photoionization calculation, what quantities to save, etc.). In the script \texttt{model.pl},  all \textit{physical} parameters characterizing the target and scattering model, except the geometries of the molecule, are specified. One or several geometries of the molecule can be given or generated in various ways in \texttt{geometry.pl}. We now describe the content of these files in detail. Note that the order of parameters given in the tables does not necessarily correspond to the order in the scripts.

\subsubsection{Nomenclature for photoionization/scattering calculation}
The UKRmol+ codes and the scripts run calculations for electron-molecule scattering or photoionization. In each case, a slightly different convention is used to define the total number of electrons in the system in the published R-matrix literature, cf.~\cite{harvey2014}. In the case of electron-molecule collisions, the number of electrons of the target molecule is $N$ so the total number of electrons is $N+1$. In photoionization, the total number of electrons is denoted as $N$ and is referred to as the initial (usually neutral) state of the molecule. The ``target" states in the photoionization case are then the final (usually positive ion) $(N-1)$-electron states of the ionized molecule. In this paper, we follow the convention for electron scattering. Regardless of the type of calculation the scripts specify in \texttt{nelectrons}   the number of electrons of the ``target" molecule (i.e. without accounting for the electron that is/will be unbound), see below.

\subsection{Directories input}
 
\begin{table}[htbp]
    \caption{Input in \texttt{dirs.pl} indicating the paths where executables, basis sets, templates, and output are located.}
    \centering
    \begin{tabular}{lll}
        \hline
        parameter &  description & format\\
         \hline
        \texttt{bin\_in} &  directory for UKRmol-in executables  & string\\
        \texttt{bin\_out} &  directory for UKRmol-out executables  & string\\
        \texttt{molpro} &  directory for Molpro executable  & string\\
        \texttt{psi4} &  directory for Psi4 executable  & string\\
        \texttt{molcas} &  directory for Molcas executable  & string\\
        \texttt{basis} &  directory where basis sets are & string   \\
        \texttt{templates} & directory where input templates are & string   \\
        \texttt{libs} & directory where libraries are & string   \\
        \texttt{output} & main directory for output in the working directory & string   \\

         \hline
    \end{tabular}
    \label{tab:dirs}
\end{table}

 Paths to the UKRmol+, Molpro, Molcas, and Psi4 executables  and to  a copy of the directories \texttt{input.templates} and \texttt{basis.sets}  are specified in \texttt{dirs.pl}. Molpro, Molcas, or Psi4 are used to generate bound molecular orbitals, but it is possible to run the UKRmol-scripts using an existing Molden file  if appropriately placed in the directory structure. Table~\ref{tab:dirs} lists all the parameters in \texttt{dirs.pl}.

\subsection{Geometry input}

The molecular geometry is provided in the \texttt{geometry.pl} file together with some additional parameters. It is possible to run the scripts for a single geometry or for a range of geometries that are explicitly specified or generated in the script. Table~\ref{tab:geom} lists first the general geometry-related  parameters set in \texttt{geometry.pl} and then those specifying a particular geometry. The latter parameters form one hash array for each given geometry and can be copied manually or automatically in the script to get an array of all geometries. The scripts provide subroutines that help the user to generate geometries by stretching particular bonds or along one or more coupled normal vibrational modes and also a subroutine to read geometries from one 
or several files in  Molden format \cite{molden}.
Examples of how to generate a range of geometries are provided with the UKRmol-scripts package in the directory \texttt{examples}. 

Parameters \texttt{start\_at\_geometry} and \texttt{stop\_at\_geometry} are only used when calculations are run for more than one geometry. It can be useful, for example, if a particular run fails for some geometries and not others and one, therefore, needs to restart it from a certain geometry, or to manually parallelize the execution on a computer cluster. To run all generated geometries set \texttt{stop\_at\_geometry} to zero.

\begin{table}[thbp]
    \caption{Geometry related input  in \texttt{geometry.pl}. }
    \centering
    \begin{tabular}{p{8.3em} p{15em} p{6.5em}}
        \hline
        parameter &  description & format\\
         \hline
        \multicolumn{3}{l}{\emph{General parameters}} \\
            \hline
        \texttt{suffix} &  added to output directory name to distinguish runs & string\\
        \texttt{geometry\_labels} & labels used on the first line of output files & string\\    
        \texttt{correct\_cm} & translate geometry so the center of mass and origin of coordinates coincide & 1: yes, 0: no\\
        \texttt{length\_unit} &  units for the coordinates &  \mbox{0: a$_0$,} 1:~angstrom\\
        \texttt{start\_at\_geometry} & only run for geometries number $>$n & integer n \\
        \texttt{stop\_at\_geometry} & only run for geometries number  $<$n & integer n\\[0.5em]
    \hline
        \multicolumn{3}{l}{\emph{Specific geometry parameters}} \\
            \hline
        \texttt{description} &  identifies each geometry in output files & string   \\
        \texttt{gnuplot\_desc} & identifies each geometry in plot files & string   \\
        \texttt{atoms} & each atom is specified by its symbol and its Cartesian coordinates in a single array, e.g. as  in \texttt{["H", 0.0, 0.0, 0.0]} &  array of arrays \\
        \hline
    \end{tabular}
    \label{tab:geom}
\end{table}

\subsection{Configuration input}

The file \texttt{config.pl} enables the user to specify runtime options of the calculation. Table~\ref{tab:basic-config} lists the basic parameters of the run. The user should modify them according to which quantum chemistry codes are available, what type of calculations to perform (photoionization, RMT input, electron/positron scattering, or target structure only), what output data to save in named files, etc.

By default, all files given on the command line after \texttt{perl main.pl} are saved into directory \texttt{scripts\_backup}; \texttt{add\_files\_to\_backup} enables the user to choose additional files to be saved. 

The scripts can generate molecular orbitals either using the Molpro or Psi4 packages. The user must set either \texttt{molpro}, or \texttt{psi4} to 1 and the other to 0. Similarly, \texttt{photoionization} and \texttt{rmt\_interface} are also mutually exclusive. However, \texttt{scattering} needs to be set to 1 even when photoionization or RMT input calculations are run.

\begin{table}[htbp]
    \caption{Basic configuration parameters in \texttt{config.pl}.}
    \centering
    \begin{tabular}{p{9em} p{15em} p{6.5em}}
        \hline
        parameter &  description & format\\
         \hline
        \texttt{suffix} &  name to distinguish different runs  & string\\
        \texttt{print\_info} & where runtime messages are printed, possible values are \emph{file}, \emph{screen}, or \emph{none} & string \\
        \texttt{add\_files\_to\_backup} & additional files to back up & array of strings \\
        \texttt{molpro} & whether Molpro will be run & 0: no, 1: yes\\
        \texttt{psi4} & whether Psi4 will be run & 0: no, 1: yes\\
        \texttt{scattering} & run the whole calculation or only the target part & 0: target only, 1: all \\
        \texttt{photoionization} & calculate photoionization cross sections & 0: no, 1: yes\\
        \texttt{rmt\_interface} & generate RMT input & 0: no, 1: yes\\
        \texttt{skip\_radden} & skip using RADDEN to calculate radial densities  & 0: no, 1: yes\\
        \texttt{skip\_time\_delay} & skip calculating time-delays  & 0: no, 1: yes\\
        \texttt{gather\_data} & gather data in a single directory & 0: no, 1: yes\\
        \texttt{clean} & remove intermediate files after run except integrals file & 0: no, 1: yes \\
        \texttt{remove\_moints} & remove integrals file after run & 0: no, 1: yes\\
        \texttt{save\_eigenph} & save eigenphase sums in named files  & 0: no, 1: yes\\
        \texttt{save\_xsec} & save cross sections in named files & 0: no, 1: yes\\
        \texttt{save\_channels} & keep channel data in named files & 0: no, 1: yes\\
        \texttt{save\_rmat\_amp} & keep raw boundary amplitudes in named files & 0: no, 1: yes\\
        \texttt{save\_Kmatrix} & keep K-matrices in named files & 0: no, 1: yes\\
        \texttt{save\_Tmatrix} & keep T-matrices in named files  & 0: no, 1: yes\\
        \texttt{keep\_inputs} &  keep input files & 0: no, 1: yes\\
        \texttt{keep\_outputs} &  keep output files  & 0: no, 1: yes\\
        \texttt{bound} & run bound state calculation   & 0: no, 1: yes\\
        \hline
        \end{tabular}
    \label{tab:basic-config}
\end{table}

Computation of radial densities using \texttt{RADDEN} and time delays is by default turned off, i.e. parameters \texttt{skip\_radden} and \texttt{skip\_time\_delay} are set to 1. Instead of \texttt{RADDEN} it is much more efficient to compute the radial densities directly as part of the integral calculation done by \texttt{SCATCI\_INTEGRALS}. This is readily achieved by setting the variable \texttt{calc\_radial\_densities = .true.} in the input template file \texttt{scatci\_integrals.inp}. The radial density for each orbital is then output to a separate file in the root directory for the calculation. This information is helpful to determine the appropriate  R-matrix radius. When \texttt{gather\_data} is set to 1, data like cross sections, eigenphase sums, etc. for different geometries are collected in files that are mainly placed in the \texttt{collected\_scattering\_data} directory. The  subsequent parameters in Table~\ref{tab:basic-config} can be used to control the deletion of temporary files created during the run and the saving of various output files.

\begin{table}[htbp]
     \caption{Expert configuration parameters for parallel execution and evaluation of integrals in \texttt{config.pl}.}
    \centering
    \begin{tabular}{p{8.3em} p{15em} p{6.5em}}
        \hline
        parameter &  description & format\\
        \hline
        \multicolumn{3}{l}{\emph{Options for parallel execution}} \\
          \hline
        \texttt{parallel\_geom} & number of geometries run in parallel & integer\\
        \texttt{parallel\_symm} & number of symmetries run in parallel & integer\\
        \texttt{mpi\_integrals} & MPI launcher command for \texttt{SCATCI\_INTEGRALS} & string\\ 
        \texttt{mpi\_scatci} & \texttt{MPI-SCATCI} launcher command & string\\
        \texttt{mpi\_rsolve} & \texttt{MPI\_RSOLVE} launcher command  & string\\[0.5em]
          \hline
        \multicolumn{3}{l}{\emph{Runtime options for \texttt{SCATCI\_INTEGRALS}}} \\
          \hline
        \texttt{buffer\_size} & size of temporary arrays for integral transformation & integer \\
        \texttt{delta\_r1} & length in $a_0$ of elementary radial quadrature needed for evaluation of mixed BTO/GTO integrals & real\\
        \texttt{transform\_alg} & integral transformation algorithm, 0: auto, 1: sparse, $>1$: dense & integer \\
        \hline
    \end{tabular}
    \label{tab:expert-config1}
\end{table}

The last parameter in Table~\ref{tab:basic-config} \texttt{bound} can be used to set up calculations when the energies of possible bound states of the total  system are needed, without performing scattering calculations. In this case, all integrals are evaluated over the whole radial range, that is, avoiding cutting off the 'tails' beyond the $R$-matrix radius, as done for inner region R-matrix calculations.

\begin{table}[htbp]
     \caption{Other expert configuration parameters in \texttt{config.pl}.}
    \centering
    \begin{tabular}{p{8.3em} p{15em} p{6.5em}}
        \hline
        parameter &  description & format\\
        \hline
        \texttt{use\_templates} & generate input files from templates or rerun calculations with user-modified input files & 0: no, 1: yes\\
        \texttt{use\_cdenprop} & force use of \texttt{CDENPROP} for target calculation & 0: no, 1: yes\\
        \texttt{ight} and \texttt{ighs} & diagonalization method for the target and all-electron calculation, respectively; $-1$:~Arpack, 0:~Davidson, 1:~Givens-Householder, 2:~automatic & integer \\
        \texttt{use\_saved\_ramps} & use amplitudes and channel data saved from a previous run & 0: no, 1: yes\\
        \texttt{run\_\{eigenp}, \texttt{tmatrx}, \texttt{ixsecs}, \texttt{reson}, \texttt{time\_delay}\} & selection of which (scattering) outer region programs to run; all can be typically switched off for photoionization runs and RMT data production & 0: no, 1: yes \\
        \texttt{dipelm\_smooth} & whether to smooth dipole moments, 0:raw data, 1: smoothed data, 2: both & integer \\
        \texttt{only} & which subset of programs to run, list of programs given in one string separated by a pipe & string \\       
        \hline
    \end{tabular}
    \label{tab:expert-config2}
\end{table}

In addition to the basic runtime options explained above, there are expert options in \texttt{config.pl}. First, in Table~\ref{tab:expert-config1}, the options for parallel execution and evaluation of integrals by the code \texttt{SCATCI\_INTEGRALS} are listed. Second, in Table~\ref{tab:expert-config2}, other expert options are summarized.

The user can decide whether to run two or more geometries and/or symmetries in parallel and whether to execute parallel versions of some programs. In the case of \texttt{mpi\_integrals}, \texttt{mpi\_scatci} and \texttt{mpi\_rsolve}, the MPI launcher command should be given explicitly, for example: \texttt{mpirun -np 32}. When these are left empty, \texttt{SCATCI\_INTEGRALS} is run serially and the older, serial versions of SCATCI and RSOLVE are used.

When BTO or mixed BTO/GTO continuum basis sets are used, the parameter \texttt{transform\_alg} should be set to 1 as using these basis sets results in large sparse matrices.

Expert options in Table~\ref{tab:expert-config2} are provided for  users who need to run calculations in a special setting or to run particular codes only. These options are briefly described in the table and more details can be found in \texttt{config.pl}. They should be used with care.

\subsection{Model input}

Settings in \texttt{model.pl} describe the physical model used for a particular calculation and are the main input required from the user. Since there are many parameters in \texttt{model.pl} it is  convenient to group them as follows:
\begin{itemize}
    \item general information about the target molecule,
    \item units to be used for outputs,
    \item details of molecular orbitals and their spaces, target states, scattering model, and total space-spin symmetries,
    \item settings for multiple active space approach,
    \item details of the pseudo-continuum orbitals,
    \item details of the continuum basis set,
    \item settings for the outer region calculation.
\end{itemize}
In addition to the groups above, some other parameters can be set. Each group of parameters is explained in detail in this section except the settings for the multiple active space approach which are described separately in Section~\ref{sec:MAS}. Most of the parameters are required for all types of calculations (photoionization, electron scattering, positron scattering, or the generation of input for RMT) but some are  only needed for some of them. When this is the case, it is stated explicitly.

Several of the parameters in \texttt{model.pl}, such as the number of orbitals or target states used, are entered as arrays of integers, where each integer corresponds to an irreducible representation of the point group of the target molecule, chosen by the user.   The position of each number in the array indicates the irreducible representation  (or simply the symmetry) with the  order as shown in Table~\ref{tab:irreps}. Since UKRmol+ works only with the basic Abelian groups listed in the table, the maximum number of symmetries is 8. When changing the point group to one with fewer symmetries (irreducible representations), it is not necessary to delete redundant elements as the scripts use only those elements up to the actual number of irreducible representations. It is recommended to set  these redundant elements to zero anyway to avoid confusion.

\begin{table}[htbp]
    \caption{Available point groups and order of their irreducible representations.}
    \centering
    \begin{tabular}{ll}
        \hline
        Point Group & Order of irreducible representations \\
        \hline
D$_{2h}$ &   A$_g$, B$_{3u}$, B$_{2u}$, B$_{1g}$, B$_{1u}$, B$_{2g}$, B$_{3g}$, A$_u$ \\
  C$_{2v}$ &   A$_1$, B$_1$, B$_2$, A$_2$ \\
  C$_{2h}$ &   A$_g$, A$_u$, B$_u$, B$_g$ \\
  D$_2$ &   A, B$_3$, B$_2$, B$_1$ \\
  C$_2$ &    A, B \\
  C$_s$ &   A$^{\prime}$, A$^{\prime\prime}$,  \\
  C$_i$ &   A$_g$, A$_u$ \\
  C$_1$ &   A \\
  	\hline
    \end{tabular}
    \label{tab:irreps}
\end{table}

\subsubsection*{General information about the target molecule}

Table~\ref{tab:input-molec} summarizes the basic information regarding the molecule that should be given in \texttt{model.pl}. We note that \texttt{atoms} should include all the atoms in the molecule, so for example, for H$_2$O this will be: 
\begin{verbatim}
'atoms',        ["O", "H", "H"], 
\end{verbatim}
The number of electrons of the target set in \texttt{nelectrons} can differ from the number for the neutral molecule. The charge of the target is then determined by the scripts from this parameter and \texttt{atoms}. The point groups that can be specified in  \texttt{symmetry} are listed in Table~\ref{tab:irreps}. Note that the scripts do not check the consistency of this option with the geometry settings in \texttt{geometry.pl} so the user should ensure that they are consistent.

\begin{table}[htbp]
    \caption{General information about the target molecule in \texttt{model.pl}. }
    \centering
     \begin{tabular}{p{5em} p{17em} p{7em}}
        \hline
        parameter &  description & format\\
         \hline
        \texttt{molecule} & name of the molecule, e.g. its chemical formula, used for directory names etc. & string \\
        \texttt{atoms} & all the atoms in the molecule given by their standard symbols in one array & array of strings \\
        \texttt{nelectrons} & number of target electrons & integer \\
        \texttt{symmetry} & point group of the molecule & string\\
         \hline
    \end{tabular}
    \label{tab:input-molec}
\end{table}

\subsubsection*{Units to be used for outputs}
Units for the output produced in the calculations, that are set in \texttt{model.pl}, are summarized in~Table~\ref{tab:input-units}.

\begin{table}[htbp]
    \caption{Units to be used for outputs in \texttt{model.pl}.}
    \centering
     \begin{tabular}{p{5em} p{17em} p{7em}}
        \hline
        parameter &  description & format\\
         \hline
        \texttt{r\_unit} & units of length & integer \\
        & 0: a$_0$, 1: angstrom & \\
        \texttt{e\_unit} & energy units & integer \\ 
        & 0: hartree, 1: Rydberg, 2: eV & \\
        \texttt{x\_unit} & units for the cross sections & integer \\
        & 0: a$_0^2$, 1: angstrom$^2$ & \\
         \hline
    \end{tabular}
    \label{tab:input-units}
\end{table}

\subsubsection*{Details of molecular orbitals and their spaces, target states, scattering model, and total space-spin symmetries}

The parameters that define the type of molecular orbitals to be generated and how these are going to be split into the different frozen,  active, and virtual  spaces determine a large part of the physics that will be modelled in the calculation. Table~\ref{tab:input-orbs} summarizes them. Note that the term \emph{frozen} is used for both closed orbitals in SCF calculations and frozen orbitals in CI calculations in \texttt{model.pl} and thus the number of these orbitals is the same in basic models. More flexible models can be set up using the MAS approach, see Section~\ref{sec:MAS}. 

\begin{table}[htbp]
    \caption{Details of the molecular orbital construction,  orbital spaces,  scattering model, and states in close-coupling expansion in \texttt{model.pl}. }
    \centering
     \begin{tabular}{p{9em} p{17em} p{6em}}
        \hline
        parameter &  description & format\\
         \hline
        \texttt{basis} & name of the basis set & string \\
        \texttt{orbitals} & which type of orbitals are generated & string \\
        & \emph{HF}: Hartree-Fock, \emph{natural}: CASSCF or MASSCF & \\
        \texttt{charge\_of} & use orbitals of \emph{target} or all-electron system (\emph{scattering}), only used for photoionization or RMT & string \\
        \texttt{select\_orb\_by} & which ordering of orbitals to use & string \\
        & \emph{molden} or \emph{energy} & \\
        \texttt{ncasscf\_states} & number of target states of each  spin and space symmetry to be averaged if a SA-CASSCF calculation is run &  hash array of integer arrays \\
        \texttt{model} & scattering model & string \\
        & \emph{SE}, \emph{SEP}, \emph{CHF-A}, \emph{CHF-B}, \emph{CAS-A}, \emph{CAS-B}, or \emph{CAS-C} & \\
        \texttt{nfrozen} & number of frozen (closed) target orbitals & integer \\
        \texttt{nactive} & number of active target orbitals & integer \\
        \texttt{nvirtual} & number of virtual orbitals used to build L$^2$ functions & integer \\
        \texttt{nreference} & number of reference orbitals & integer \\
        \texttt{frozen\_orbs} & frozen orbitals for each symmetry & integer array \\
        \texttt{active\_orbs} &  active orbitals for each symmetry & integer array \\
        \texttt{virtual\_orbs} & virtual orbitals for each symmetry & integer array \\
        \texttt{reference\_orbs} & reference orbitals for each symmetry & integer array \\
        \texttt{ntarget\_states} & number of target states of each total spin and symmetry for which energy and permanent and transition moments are calculated & hash array of integer arrays \\
        \texttt{ntarget\_states\_used} & number of target states included in the close-coupling expansion & integer \\
        \texttt{scattering\_states} & which total symmetries to run  the scattering calculation for & hash array of integer arrays \\
         \hline
    \end{tabular}
    \label{tab:input-orbs}
\end{table}


The basis sets that can currently be indicated in \texttt{basis} and   used are those provided with the package (in the \texttt{basis.sets} directory). The user can create the file for other standard (or tailor-made) basis sets. 

The parameter \texttt{charge\_of} is only used for photoionization or RMT input calculations and indicates whether orbitals for the target (usually a cation) or the total system are to be used.

As the scripts always use the first $n$ orbitals (for all symmetries or each symmetry) requested by the user it is important to select an appropriate ordering using \texttt{select\_orb\_by}. When this is set to \emph{molden}, the orbitals are used in the order in which they are saved in the Molden file. If set to \emph{energy}, then the orbitals are used in energy order. This parameter is irrelevant if Hartree-Fock orbitals, which are always ordered by energy, are used.

The UKRmol-scripts will run either Molpro, Psi4 or Molcas to generate bound molecular orbitals, either using the Hartree-Fock (HF) method, the state-averaged complete active space self-consistent field (SA-CASSCF) approach, or the state-averaged multiple active space self-consistent field (SA-MASSCF) approach (which comes in two variants SA-ORMASSCF and SA-GASSCF). For details on how to set up the last approach, which is only available when using Molpro or Molcas, see Section~\ref{sec:MAS}. Table~\ref{tab:QCMethods} summarises which quantum chemistry methods can be used with each quantum chemistry code.

\begin{table}[htbp]
    \caption{Quantum chemistry methods available in the scripts; 'irrep' stands for irreducible representation. }
    \centering
    \begin{tabular}{p{17em} p{3.5em} p{3.5em} p{3.5em}}
        \hline
        Method &  Molpro & Psi4 & Molcas \\
        & 2021.2 & 1.7 & 8.4 \\
         \hline
        HF & \checkmark & \checkmark & \checkmark \\
        CASSCF & \checkmark & \checkmark & \checkmark \\
        ORMASSCF & \checkmark &  &  \\  
        GASSCF & & & \checkmark \\
        State Averaging: Single irrep & \checkmark & \checkmark & \checkmark \\
        State Averaging: Multiple irreps & \checkmark & &  \\
        State Averaging: Multiple spins & \checkmark$^*$& &  \\
        $^*$Not when using ORMASSCF \\
         \hline
    \end{tabular}
    \label{tab:QCMethods}
\end{table}

When the SA-CASSCF approach is chosen, the frozen and active spaces as well as the states to be averaged, need to be selected.
The latter is done using \texttt{ncasscf\_states} to indicate how many states of each space-spin symmetry are averaged; note that currently the scripts only allow for state-averaging with equal weights for all the states. The frozen and active spaces will  also be used to generate the target and total (all-electron) wave functions in the UKRmol+ suite, with the addition of some virtual orbitals.

The parameter \texttt{model} defines how the total wavefunctions are going to be built. The scattering models used correspond to the standard approximation levels  in low-energy electron and positron scattering \cite{tennyson,UKRMOL+}: static or static-exchange (SE), for positrons and electrons respectively, static-exchange plus polarization (SEP), or close-coupling (CAS) where the choice of A, B or C leads to different types of L$^2$ functions. Additional models CHF-A and CHF-B are described in Sec.~\ref{sec:selectedmodels}. 

The number of frozen, active, and virtual orbitals to be used is set by \texttt{nfrozen}, \texttt{nactive}, and \texttt{nvirtual} parameters. By default, they will be selected in the order indicated by \texttt{select\_orb\_by}.
However, it is possible to request a specific number of orbitals of each symmetry by using \texttt{frozen\_orbs}, \texttt{active\_orbs}, and \texttt{virtual\_orbs}. The total number of chosen orbitals of each type must be equal to \texttt{nfrozen}, \texttt{nactive}, and \texttt{nvirtual}.

The parameters for frozen and active orbitals are not used in SE calculations. For SEP calculations, the $L^2$ functions only involve single excitations from the active space into the virtual space. For close-coupling calculations, the active orbitals are used in the generation of the configurations for the target and the total wave functions. The virtual orbitals, in all cases, are only used to build the L$^2$ functions.

Parameters \texttt{nreference} and \texttt{reference\_orbs} do not impact the physics described. They are required as  input to program \texttt{CONGEN} that builds the configuration state functions for target and total wave functions relative to a reference determinant. Providing  \texttt{nreference} and \texttt{reference\_orbs} is particularly useful in the case of electron-rich targets and, when possible, it is suggested that all orbitals occupied in the ground state of the target plus a few additional orbitals of each symmetry are used.

For close-coupling calculations, one must set \texttt{ntarget\_states\_used} to the number of target states to be included in the close-coupling expansion. The program will select the \texttt{ntarget\_states\_used} lowest states out of the total number of states for which target properties are evaluated. Using \texttt{ntarget\_states} the user can request that these properties are calculated for a specific number of states of each possible spin-space symmetry. The spin needs to be stated explicitly (so, for a target with an even number of electrons, 'singlet' and 'triplet' are usually specified) and then the number of states of each symmetry.

It is usually  helpful to set \texttt{ntarget\_states} such that the total number of states for which properties are calculated is larger than \texttt{ntarget\_states\_used}. This enables the user to check and ensure that no states are skipped in the close-coupling expansion as this could lead to non-physical features in the evaluated physical quantities \cite{tennyson}.

In UKRmol+, once the integrals are evaluated, the inner region calculations are run separately for each space-spin symmetry of the total system. Sometimes (for example, if a specific resonance wants to be investigated), it is desirable to run calculations for a reduced number of space-spin symmetries and  \texttt{scattering\_states} can be used for this purpose. As with \texttt{ntarget\_states}, the spin need to be stated explicitly (so, for a total system with an odd number of electrons, 'doublet' and 'quartet', the latter for scattering calculations, are usually specified) and then a zero or one to indicate if that space-spin symmetry is required.

\subsubsection*{Details of the pseudo-continuum orbitals}

The R-matrix with pseudostates method (RMPS) \cite{bhs96} is a generalization of the close-coupling method that uses an extra set of target orbitals, known as pseudo-continuum orbitals (PCO), to provide a representation of the discretized continuum in the inner region. Table~\ref{tab:input-pco} lists the parameters in \texttt{model.pl} associated with the use of PCos and pseudostates.

The molecular implementation of this method uses even-tempered GTOs \cite{GoT2004} where the exponents are determined by:
\begin{equation}\label{PCO_exp}
\alpha_i= \alpha_0 \times \beta^{i-1},\quad i=1,\ldots,N_e.
\end{equation}
Values of $\alpha_0$, $\beta$, and $N_e$ can differ for each partial wave, thus parameters \texttt{PCO\_alpha0}, \texttt{PCO\_beta}, and \texttt{num\_PCOs} are actually arrays of values. If these orbitals are not required, set \texttt{use\_PCO} to zero and there is no need to modify any of the other parameters.

\begin{table}[htbp]
    \caption{Details of the pseudo-continuum orbitals in \texttt{model.pl}. }
    \centering
    \begin{tabular}{p{7em} p{17em} p{6em}}
        \hline
        parameter &  description & format \\
         \hline
        \texttt{use\_PCO} & switch to include PCO basis & 0: no; 1: yes \\
        \texttt{reduce\_PCO\_CAS} & selects which configurations to include & 0: no; 1: yes \\
        \texttt{maxl\_PCO} & highest partial wave used in the PCO GTO basis & integer \\
        \texttt{PCO\_alpha0} & $\alpha_0$ for PCO generation & real array \\
        \texttt{PCO\_beta} & $\beta$ for PCO generation & real array \\
        \texttt{num\_PCOs} & number $N_e$ of PCO generated & integer array \\
        \texttt{PCO\_gto\_thrs} &  threshold for allowed continuum GTO exponents & real array \\
        \texttt{PCO\_delthres} & deletion threshold for orthogonalization & real array \\
         \hline
    \end{tabular}
    \label{tab:input-pco}
\end{table}

There is no restriction in principle on the maximum partial wave to be included. However, due to the computational cost of this additional basis set, this is normally kept as low as or lower than the maximum partial wave for the continuum orbitals.  To avoid severe linear dependence, \texttt{PCO\_gto\_thrs} ensures that the exponents of continuum and PCO GTOs are not too similar:  continuum basis functions with exponents equal or bigger than \texttt{PCO\_gto\_thrs} are removed from the basis. If set to -1, then a default value of $\texttt{PCO\_alpha0} \times \left(\texttt{PCO\_beta} - 1 \right)$ is used as threshold.

 When PCOs are used, these are orthogonalized to the target bound orbitals first and then the continuum orbitals are orthogonalized to the set of target orbitals + PCOs. The parameter \texttt{PCO\_delthres} contains deletion thresholds for each symmetry that fulfills the same role as \texttt{delthresh} for the continuum (see below and Table~\ref{tab:input-cont}).

When PCOs are included in the calculation, target and total configurations involving single (and in the latter also double) occupancy of the PCOs are used. For models where the target  wave functions  are generated using an active space, the configurations that include PCOs can be built in two ways, controlled by the parameter \texttt{reduce\_PCO\_CAS}. When \texttt{reduce\_PCO\_CAS} equals 1, the configurations
involving PCOs are generated by promoting an electron from an orbital occupied in the ground state (Hartree-Fock) configuration to a PCO. 
Alternatively, PCO configurations are generated by promoting an electron from the  active space into a PCO. The latter often leads to a significantly larger number of configurations. For this reason, it is usual to set \texttt{reduce\_PCO\_CAS} to 1.

\subsubsection*{Details of the continuum basis set}

The continuum orbitals can be expanded in a set of GTOs with predetermined exponents (these are read from the \texttt{basis.sets} directory), a set of BTOs that are defined via input parameters, or a mixed GTO-BTO basis set.

\begin{table}[htbp]
    \caption{Parameters defining continuum orbitals in \texttt{model.pl}. }
    \centering
    \begin{tabular}{p{8em} p{16em} p{6em}}
        \hline
        parameter & description & format\\
         \hline
        \texttt{use\_GTO} & switch to use GTOs & 0: no; 1: yes\\
        \texttt{radius\_GTO} & which GTO continuum basis to use & integer\\
        \texttt{maxl\_GTO} & highest partial wave for GTOs & integer\\
        \texttt{use\_BTO} & switch to use BTOs & 0: no; 1: yes\\
         \texttt{start\_BTO} & radius where the BTOs start & integer \\
        \texttt{rmatrix\_radius} & R-matrix radius, where the BTOs end & real\\
        \texttt{order\_BTO} & order of the BTOs & integer \\
        \texttt{no\_of\_BTO} &  number of BTOs  & integer \\
        \texttt{maxl\_BTO} & highest partial wave used for BTOs & integer\\
        \texttt{delthres} & deletion thresholds for the continuum orthogonalization, one per symmetry & real array \\ 
        \texttt{maxl\_legendre\_1el} & maximum $L$ in Legendre expansion for nuclear attraction integrals & integer \\
        \texttt{maxl\_legendre\_2el} & maximum $L$ in Legendre expansion for nuclear 2-electron integrals & integer \\
       \hline
    \end{tabular}
    \label{tab:input-cont}
\end{table}

The parameter \texttt{radius\_GTO} determines which GTO continuum basis will be used.  Continuum GTO bases are provided in the \texttt{basis.sets} directory for the following radii: 4, 6, 10, 13, 15, and 18~$a_0$ for neutral targets and 10 and 13~$a_0$ for positively charged ones. Since the exponents of the GTO continuum basis must be optimized for a specific radius for each partial wave, the maximum angular momentum available for each combination of radius and charge state may differ. If a different value is set for \texttt{radius\_GTO}, the user will need to generate the continuum GTO basis set using GTOBAS \cite{gtobas}, a program from the UKRmol+ suite. 

If no BTOs are included in the calculation,  \texttt{rmatrix\_radius} must be set equal to \texttt{radius\_GTO}. If a mixed GTO-BTO basis is used, the parameter \texttt{start\_BTO} must be chosen to be less than \texttt{radius\_GTO} and a standard choice is $\texttt{start\_BTO}< \texttt{radius\_GTO}-1\ a_0$.

Parameters \texttt{maxl\_legendre\_1el} and \texttt{maxl\_legendre\_2el} control the numerical evaluation of the nuclear one- and two-electron integrals involving BTOs in \texttt{SCATCI\_INTEGRALS}; for details, see \cite{UKRMOL+}.

Since UKRmol+ works with orthogonal orbitals, the continuum basis sets must be orthogonalized both to the target orbitals that will be used and among themselves. The latter is done using symmetric orthogonalization and this requires setting a threshold \cite{UKRMOL+}:  continuum orbitals with eigenvalues of the overlap matrix smaller than this threshold, given by \texttt{delthres}, are deleted. Different thresholds can be given for different orbital symmetries, although usually the same value is used for all.

\subsubsection*{Settings for the outer region calculation}

The outer region calculation will produce one or more photoionization or scattering quantities (cross sections, asymmetry parameters, K-matrices, etc.) for a grid of kinetic energies of the unbound particle. The grid is defined as one or more energy subranges $j$ containing $N_j$ energies given by:
\begin{equation}
E_i= E^j_{\mathrm{inc}}(1) + E^j_{\mathrm{inc}}(2)\times i\ \mathrm{for}\ i=1,...,N_j 
\end{equation}
where the starting energy of the subrange $E^j_{\mathrm{inc}}(1)$, increment of energy $E^j_{\mathrm{inc}}(2)$ and $N_j$ are set in \texttt{model.pl} (see Table~\ref{tab:input-outer}).

\begin{table}[htbp]
    \caption{Parameters for the outer region calculation in \texttt{model.pl}. }
    \centering
    \begin{tabular}{p{7em} p{18em} p{4.5em}}
        \hline
        parameter &  description & format\\
         \hline
        \texttt{raf} & propagation radius & real\\
        \texttt{max\_multipole} & maximum multipole retained in expansion of long-range potential & integer $\leq$2 \\
        \texttt{nescat} & number of energies $N_j$  in each subrange, string of integers separated by commas & integer array \\
        \texttt{einc} & initial energy $E^j_{\mathrm{inc}}(1)$ and energy increment $E^j_{\mathrm{inc}}(2)$ for each subrange,  pairs of real numbers separated by commas & real array \\
        \texttt{maxi} & highest initial state for which cross sections, etc. are calculated &  integer\\
        \texttt{maxf} & highest final state for which cross sections, etc. are calculated & integer\\
        \hline
    \end{tabular}
    \label{tab:input-outer}
\end{table}

In the case of scattering calculations for neutral targets, the propagation radius \texttt{raf} is usually set to values larger than 50~$a_0$. The bigger the value, the longer the outer region calculations. However, \texttt{raf}  of 100~$a_0$ can be necessary, for example, when channels are energetically close. 

By default, the target calculation will produce dipole and quadrupole permanent and transition moments for all  target states considered. \texttt{max\_multipole} enables the user to decide whether any of these will be used to model the interaction of the target and the unbound particle in the outer region. In scattering calculations, it is customary to set it to 2. If set to zero, this switches off the interaction in the outer region. For photoionization calculations, the outer region calculation is set up differently, see Section~\ref{sec:photo}.

\subsection{General description of output}

The output of one run of the scripts varies according to settings. In general, the scripts will create a directory for a specific run (either named  automatically according to the  model parameters and the \texttt{suffix} given in \texttt{geometry.pl} and \texttt{config.pl}, or  specified explicitly by the user in \texttt{model.pl} using \texttt{directory}). The following subdirectories are created during the run:
\begin{description}
   \item[\texttt{scripts\_backup}:] subdirectory where all the scripts (\texttt{.pl} files) used for the particular run are copied for later reference,
   \item[\texttt{logs}:]  subdirectory where all log files are stored; a \texttt{main.log} file and one \texttt{geom<n>.log} file for each geometry are created,
   \item[\texttt{geom<n>}:]  subdirectory where all input and output files of each program in the UKRmol+ suite for each  geometry  $n$ are stored,
   \item[\texttt{collected\_scattering\_data}:]  subdirectory where the files with generated scattering  data, such as eigenphase sums, T-matrices, and cross sections, are copied from subdirectories \texttt{geom<n>} for simpler handling and plotting.
   \end{description}
In addition, the following files are created, usefully collating data for several geometries:
      \begin{description}
    \item[\texttt{geometries} and \texttt{geometries.molden}:]  text files containing descriptions of all used molecular geometries, a simple sequence of numbers $n$ is assigned to the geometries, which are used for naming of output files and a subdirectory for each specific geometry,
   \item[\texttt{target.energies}:]  text file with energies of the target electronic states included in the calculations for all geometries,
   \item[\texttt{target.dipole.moments}:]  text file with the permanent dipole moment of the ground target electronic state for all geometries,
   \item[\texttt{Rmatrix.energies}:]  text file with several of the lowest-energy R-matrix poles for all geometries,
   \item[\texttt{resonance.positions.and.widths}:]  text file with  resonance positions and widths for all geometries for those resonances identified and characterised by the program \texttt{RESON} \cite{RESON}.
\end{description}

The scripts also create a few files (with the extension \texttt{.gp}) for easier plotting of target energies, eigenphase sums, and cross sections using the Gnuplot software. These files are useful to compare results for several geometries.

\section{Running scattering calculations}\label{sec:example_scat}

The workflow for scattering calculations is shown in Figs.~4 and 5 of the UKRmol+ paper \cite{UKRMOL+}.

\subsection{Input}

For all types of calculations, the user should choose the geometry or geometries to be run. The ground state configuration of the target is also required. When the calculations include several target states, a number of tests to determine the most appropriate basis set, active space and/or state averaging are needed. For this step, it is useful to set \texttt{scattering}${}=0$ to reduce the computational effort. Once the best model is chosen, setting \texttt{scattering}${}=1$ ensures that all requested programs are executed.

The lepton projectile is  specified via the \texttt{positron\_flag} in \texttt{model.pl}: it should be set to 0 for electron scattering  and 1 for positron scattering. If not specified,  the default value is 0 (electron scattering).  In positron scattering, excitation of triplet target states is not allowed due to selection rules. Therefore,  triplet \texttt{ncasscf\_states} and \texttt{ntarget\_states} should all be equal to 0 for positron scattering. 


The default outer region programs run by the scripts are \texttt{SW\_INTERF} and \texttt{RSOLVE} so that K-matrices are generated. The user should then select, in the \texttt{config.pl} file (see Table~\ref{tab:expert-config2}), which other programs to execute. This will determine what output is produced.


\subsection{Outputs generated}
The same output files are generated for electron and positron scattering,  except \texttt{RESON} is not usually run for the latter so no resonance information is produced. In the case of positrons, the directory name will include "e+" (instead of "e-"). 

In addition to the output file for each program in the UKRmol+ suite, a scattering calculation will produce several output files:
\begin{itemize}
    \item In each \texttt{geom<n>} directory, files containing the contribution to the cross sections from each irreducible representation, as well as the eigenphase sum for each of these geometries.
    \item The same data will also be collected in the \texttt{collected\_scattering\_data} directory in an easy-to-plot format.
    \item If selected in the \texttt{config.pl}  script, the K and/or T-matrices will be saved into \texttt{collected\_scattering\_data}.
    \item If resonances are identified and characterized using \texttt{RESON}, a file named \texttt{resonance.positions.and.widths} with the energy and width of  the resonances detected for all geometries is generated.
\end{itemize}

\subsection{Electron scattering example}
We provide an example of a close-coupling calculation for H$_2$O. The calculation is run for four different geometries, generated automatically by stretching the OH bond lengths symmetrically. It uses the C$_s$ point group.

The calculation employs the cc-pVTZ basis set and CASSCF orbitals generated by Molpro for the ground state. These orbitals have been saved in files named \texttt{h2o.molden}, one for each symmetry, to enable the user to run a UKRmol+ calculation without the need for Molpro to demonstrate the use of the \texttt{only} option of the scripts (see Table~\ref{tab:expert-config2}).

The 1a$^{\prime}$ orbital is kept frozen, the active space comprises five a$^{\prime}$  and one a$^{\prime\prime}$ orbitals. Five target states are included in the scattering calculation. The continuum is GTO only and includes partial waves up to $l=4$. The R-matrix radius used is 10~$a_0$ and the propagation is carried out to 70~$a_0$.

The calculation is set up to run serially and to save cross sections and eigenphase sums only, collected in the \texttt{collected\_scattering\_data} directory. The following outer region programs are run: \texttt{EIGENP}, \texttt{TMATRX}, \texttt{IXSECS} and \texttt{RESON}. Resonance information is collected in the usual file.

\subsection{Positron scattering example}
\label{}

Two examples are provided for positron scattering calculations. The simpler one uses a similar model to that used for electron scattering from H$_2$O, but it set to run for just one geometry. Again, the Molden file with the orbitals is provided. 

The second example shows the use of pseudocontinuum orbitals in a positron calculation. In this case, the target is H$_2$ and the D$_{2h}$ point group is used. No Molden file is provided for this example, so a quantum chemistry program is needed to run this test. The 6-311G** basis set is used to generate CASSCF orbitals for the ground state. The active space contains 9 orbitals, of the following symmetry: three a$_g$, one b$_{3u}$, one b$_{2u}$, two b$_{1u}$, one b$_{2g}$ and one b$_{3g}$. In addition, PCOs are included in the calculation: these are generated using $\alpha_0$=0.17 and $\beta$=1.4 for $l$=0,1,2. 

15 singlet target states are included in the close-coupling, corresponding to the ground state and single excitations into the PCOs. The R-matrix radius is set to 13~a$_0$ and a GTO continuum basis with  partial waves up to $l=4$ employed; again, the propagation radius is set to 70~$a_0$.

\section{Running photoionization calculations}\label{sec:photo}

For calculations of (perturbative) single-photon ionization, the scripts implement the workflow depicted in Fig. 6 of the UKRmol+ paper~\cite{UKRMOL+}. The calculation formally proceeds as if for the case of electron-molecule scattering where the ``target" molecule is represented by the final states of the photoionized system (typically a positive ion). In both cases a continuum wavefunction is generated but for different asymptotic boundary conditions (outgoing wave in case of scattering and incoming wave in case of photoionization).

In addition to the calculation of the continuum wave functions, the photoionization run produces dipole matrix elements between the R-matrix basis states (the eigenfunctions of the Hamiltonian for the whole system in the inner region) which are used to assemble the physical photoionization dipole matrix elements. This is done by the \texttt{CDENPROP} program~\cite{harvey2014}.

\subsection{Input}

The input parameters required to specify the molecular model are identical to the scattering calculation. In the usual case of photoionization of neutral molecules the photoelectron moves, asymptotically, in the field of the charged molecule. As a result some physical parameters of the calculation are chosen differently:
\begin{itemize}
    \item \texttt{max\_multipole} should be set to 1 since the dominant interaction in the outer region is Coulomb.
    \item For the same reason the propagation radius \texttt{raf} should be set to the R-matrix radius (i.e. no propagation is performed). This choice has practical reasons too: in photoionization backpropagation of the wavefunction from the asymptotic to the R-matrix radius would have to be performed which is subject to numerical instabilities in the case of closed channels. If long-range interactions in the outer region are important these should be effectively included by increasing the size of the R-matrix radius instead.
    \item In addition, the scripts automatically ensure that the nuclear contributions to the molecular (bound-bound) multipole transition moments are not included if \texttt{photoionization} is set to 1 (this  the \texttt{isw} option  the \texttt{DENPROP} program). The nuclear contribution to the dipole matrix elements must be excluded since the bound-bound dipole transition moments are needed in the \texttt{CDENPROP} program to compute the light-driven dipole transitions between the R-matrix basis states. As a result, the multipole matrix elements generated by \texttt{DENPROP} from the photoionization run cannot be used in the outer region to represent the long-range electron-molecule interaction. For that reason, it is important to switch off the R-matrix propagation in the outer region.
    \item The symmetry of the initial molecular state is specified in \texttt{initialsym}. By default, the initial state in the photoionization dipole matrix element is taken as the lowest R-matrix basis state of that symmetry.
    \item \texttt{first\_Ip}, in units of \texttt{e\_unit}, can be used to specify the accurate first ionization potential (Ip) of the molecule. This value is used only by the \texttt{DIPELM} program to shift the first Ip to the accurate value when computing the photoionization observables. If this value is set to zero then the calculated Ip is used.
    \item \texttt{dipelm\_smooth} controls whether the smoothed photoionization observables should be produced by smoothing of the partial wave dipole matrix elements. Typically, the photoionization of neutral molecules leads to a congested spectrum of sharp autoionization states. Consequently, smoothing of the observables may produce artifacts in the observables, particularly close to the threshold. For that reason, it is advised to always produce unsmoothed (unbiased) results too. Set \texttt{dipelm\_smooth} to 0 for unsmoothed data, 1 for smoothed data, or 2 for both.
\end{itemize}

In photoionization, depending on the symmetry of the molecule, dipole transitions from the initial state, which has a specific symmetry, may only couple to a subset of the irreducible representations representing the continuum. In this case, only a subset of the \texttt{scattering\_states} has to be generated, see Table~\ref{tab:input-orbs}.

\subsection{Outputs generated}

The output of the photoionization calculation consists of the standard orientation averaged photoionization cross sections and angular distributions ($\beta$ parameters) saved in the files with self-explanatory names starting with ``\texttt{photo\_}". In addition, the partial wave dipole matrix elements are saved on the files \texttt{pwdips-x}, \texttt{pwdips-y} and \texttt{pwdips-z}. These files are placed in the \texttt{geom<n>} directories.

\subsection{Example}

The example provided runs a close-coupling photoionization calculation for CH$_4$. A small active space of $8$ electrons distributed in $6$ orbitals is used to describe the ionic states of the molecule. The calculation is performed in the reduced D$_2$ point group. It includes four final states of the ion in the close-coupling expansion and a small Gaussian continuum with partial waves up to $l=4$. The scattering model uses a small set of additional L$^2$ functions corresponding to single excitations out of the active space to a single virtual orbital. In addition, the calculation is performed using a set of CASSCF molecular orbitals pre-generated in Molpro (saved in the file \texttt{ch4+.molden}). The reference photoionization cross sections and angular distributions are provided too.

\section{Generating input for RMT}\label{sec:rmt}

For simulations of laser-driven electronic processes in molecules, the RMT package~\cite{RMTMOL} requires a single binary file \texttt{molecular\_data} produced by the program \texttt{RMT\_INTERFACE}. The associated workflow is shown in Fig.~7 in \cite{UKRMOL+}. The most sizeable part of this file is the transition dipole moments between all pairs of R-matrix basis states. Further data, including a detailed description of outer region channels, boundary amplitudes, and various useful coupling coefficients is also included.

Molecular RMT requires boundary amplitudes evaluated not only at the boundary between the inner and outer region
but also at several equally spaced radii inside the R-matrix sphere. The number of these extra radii and their spacing are set to a reasonable default directly in the UKRmol-scripts input templates (18 and \(0.08\,a_0\) respectively). The chosen values correspond to the default setup of RMT. The defined overlap region of \(18 \times 0.08\,a_0 = 1.44\,a_0\) needs to be numerically free of the target orbitals, which may require the use of a larger R-matrix radius than the corresponding photoionization run.
Note that there is no  keyword in the scripts to change these values; if needed, they can be changed directly in the templates. 

Note that the default workflow that uses the program \texttt{RMT\_INTERFACE} has an internal limit on the molecular model size. If this is exceeded, the program will fail. When such a large model is needed, it is possible to use the \texttt{MPI-SCATCI}-based alternative UKRmol+ workflow (see Fig.~8 in \cite{UKRMOL+}), which does not suffer from such limitations. This alternative mode is enabled by setting \texttt{parallel\_symm} to ``0'' in \texttt{config.pl}.

\subsection{Input}

To generate the RMT file for a particular molecular model, one sets the options \texttt{scattering} and
\texttt{rmt\_interface} in \texttt{config.pl} to ``1''; the option \texttt{photoionization} should be set to ``0''.
Otherwise, the remaining setup should correspond to that of a photoionization calculation. The only difference with respect to a typical photoionization calculation is the number of  ``scattering'' symmetries:
in one-photon ionization sometimes only a subset of irreducible representations is required due to symmetry reasons.
However, multi-photon or strong-field processes with arbitrary polarizations couple wave functions of all irreducible representations. As a consequence, the RMT interface mode requires
setting all spin-accessible symmetries in \texttt{scattering\_states} in \texttt{model.pl} to ``1''.

\subsection{Outputs generated}

Even though the output of the scripts in this mode is similar to the photoionization run, the only
relevant output is the file \texttt{molecular\_data} present in the geometry folder.

\subsection{Example}

We provide a simple example to generate input data for RMT for a Static Exchange calculation for H$_2$ molecule with a small Gaussian-only continuum basis. The calculation is performed in D$_{2h}$ point group symmetry and uses a set of Hartree-Fock molecular orbitals pre-generated, this time, by Psi4. Dipole transition matrices between the singlet R-matrix states of all irreducible representations are generated to be used in the inner-region part of the RMT calculation and saved on the final output file \texttt{molecular\_data} along with all other auxiliary quantities as described above.

\section{Description of selected models} \label{sec:selectedmodels}
The standard SE, SEP and close-coupling models, as well as the use of pseudocontinuum orbitals and pseudostates has been described in detail elsewhere, in particular in \cite{UKRMOL+}. Here, we describe two newer approaches to modelling electron correlation and generating the configuration state functions for the target and all electron systems.

\subsection{Multiple active space (MAS) approach}\label{sec:MAS}

The multiple active space approach (MAS) is a new method for defining the 
orbital active space for UKRmol+ calculations \cite{harvey_mas}.
It is based on a generalisation of the occupation restricted multiple active space approach (ORMAS) \cite{ormasscf} and the generalised active space approach (GAS) \cite{gasscf}. The input required to set this model are summarized in Table~\ref{tab:MAS}.

\begin{table}[htbp]
    \caption{Using the MAS approach in \texttt{model.pl} }
    \centering
    \begin{tabular}{p{9em} p{17em} p{3.5em}}
        \hline
        parameter &  description & format\\
         \hline
        \texttt{use\_MASSCF} & use MASSCF for quantum chemistry & integer\\
        & 0: don't use MASSCF & \\
        & 1: ORMASSCF in Molpro & \\
        & 2: GASSCF in Molcas & \\
        \texttt{use\_MAS} & use MAS for target and scattering & integer\\
        & 0: don't use MAS & \\
        & 1: use ORMAS & \\
        & 2: use GAS & \\        
        \texttt{qchem\_MAS} & allows for a different MAS for the quantum chemistry of the target and total calculations & array\\
        \texttt{qchem\_constraints} & additional constraints on allowed CSF & function\\
        \texttt{MAS} & define the target active spaces & array\\
        \texttt{constraints} & additional constraints on allowed CSF & function \\
        \texttt{l2\_MAS} & define the $L^2$ active spaces if different from an extra electron in \texttt{MAS} & array \\
        \texttt{l2\_constraints} & additional constraints on allowed CSF & function   \\
         \hline
    \end{tabular}
    \label{tab:MAS}
\end{table}

ORMAS and GAS were developed as techniques for excluding unnecessary CSF from the configuration space generated by the standard complete active space (CAS) approach.
They work by dividing the active space into multiple subspaces and defining a maximum and minimum electron occupation for each subspace.
ORMAS defines a local minimum and maximum, while GAS defines a cumulative minimum and maximum.
It is often possible to reduce the number of CSF by an order of magnitude or more using the MAS approach while retaining chemical accuracy, both in the bound states \cite{ormasscf, gasscf} and in the scattering or photoionization observables \cite{harvey_mas}.

A second benefit of the MAS approach is that it provides a simple yet flexible
method of defining active spaces in general, allowing for sophisticated models beyond the predefined models available in the scripts.

MAS can be used both for generating the orbitals in the initial quantum chemistry
calculation (ORMASSCF in Molpro and GASSCF in Molcas, we will use MASSCF to refer
to both) and for specifying the active space in the following target and 
total wave function calculations.

A MAS subspace is defined by a triplet of elements: First is an array specifying the orbitals per symmetry in the subspace, or an integer specifying the total number of orbitals in the subspace.
If the second method is used then the orbitals are picked in order, as defined by \texttt{select\_orb\_by}, and the orbitals per symmetry are populated automatically.  
Second is an array giving the minimum and maximum occupancy.
Thirdly, a string describing the subspace, \verb|"frozen"/"closed"/"active"|, this string is optional for target and total calculations but required when doing a MASSCF quantum chemistry calculation.
For example, using the ORMAS version of the MAS approach, a 10-electron molecule, with 2 electrons in a closed orbital, and single, double, and triple excitations out of a 6 orbital subspace into a 5 orbital subspace would be written as 
\begin{verbatim}
$model{MAS} = [
    1,[2,2],"closed", 6,[5,8],"active", 5,[0,3],"active"
]
\end{verbatim}
In the GAS version, this would be
\begin{verbatim}
$model{MAS} = [
    1,[2,2],"closed", 6,[7,10],"active", 5,[10,10],"active"
]
\end{verbatim}
The parameter \texttt{constraints} allows for further constraints to be applied to the subspaces by passing an anonymous function that removes unwanted distributions of electrons across the subspaces. For example, using ORMAS, the following model, 
\begin{verbatim}
$model{MAS} = [
    5,[6,10],"active", 2,[0,4],"active", 5,[0,1],"active"
]
$model{constraints} = 
    sub {my $dist=shift; !($dist->[2] > 0 && $dist->[0] != 9);}
\end{verbatim}
allows single excitations to the third subspace only when there are 9 electrons in the first subspace, if HF orbitals are used then this is equivalent to CAS(10,7) + single excitations from the HF configuration to 5 higher-lying orbitals.
Note that when using constraints we need to set the L$^2$ space by hand, as the
scripts are not clever enough to derive \texttt{l2\_constraints} from \texttt{constraints} on their own.

If MAS is used for the target and total run but MASSCF is not used for the quantum chemistry calculation then all the variables that are usually set to define the active space for the quantum chemistry calculation still need to be set, i.e. \texttt{nfrozen}, \texttt{nactive}, \texttt{nvirtual}, and possibly \texttt{frozen\_orbs}, \texttt{active\_orbs} and \texttt{virtual\_orbs}. 

If MASSCF is used, then only \texttt{nvirtual} and possibly \texttt{virtual\_orbs} need to be set as the rest are determined by the quantum chemistry MAS. In all cases \texttt{nreference}, and possibly \texttt{reference\_orbs} still need to be set.

The MAS for the target and total calculations does not need to respect the
type of active spaces defined in the quantum chemistry calculation. E.g.
orbitals that are frozen or virtual in the quantum chemistry calculation can
be specified as active in the target and total calculation. However, 
the total number of orbitals in the target calculation cannot exceed those
defined in the quantum chemistry calculation. One thing to note is that
any virtual orbitals retained from the quantum chemistry calculation that 
are not explicitly used in the MAS for the target and total calculation 
are automatically contracted with the continuum orbitals. This means that they are formally included as part of the 1-electron continuum orbital basis.

A number of examples of \texttt{model.pl} files setting different MAS calculations are available in  the directory \texttt{examples}.

\subsection{Polarization-consistent coupled Hartree–Fock approach}\label{sec:PCCHF}

The polarization-consistent coupled Hartree–Fock (PC-CHF) approach uses a simple Hartree–Fock-like
description of the target states to model polarization and multi-channel effects in polyatomic
molecular targets. The model is constructed in a self-consistent manner meaning that all of the target
states implied by the polarization configurations are included in the total wavefunction. This model has
been used recently \cite{PCCHF} to model photoionization of three different molecules: H$_2$O,
N$_2$O and formic acid.

PC-CHF is similar to the SEP model in that it retains a simpler HF-like description of the target
states, but it couples polarization states to their implied ionic states in order to remove spurious
resonances that are observed in SEP calculations. For more details on the model and how
states are generated see Section 2.1 in \cite{PCCHF} .

The PC-CHF model is selected using the appropriate keyword in the \texttt{model} parameter of the \texttt{model.pl} file and by specifying HF orbitals in the parameter \texttt{orbitals}. 
Two variants of the PC-CHF model are available: \emph{CHF-A} and \emph{CHF-B}.
Setting \texttt{model} to \emph{CHF-A} is recommended for most use-cases. In this model, each HF-like target configuration corresponds to a single target state in the close-coupling expansion.
Whereas, in \emph{CHF-B}
the target configurations of the same spin-space symmetry are effectively producing CI expansions of the resulting target states. The latter sub-model was designed to enable PC-CHF calculations for molecules with degenerate target molecular orbitals of different symmetries that lead to degenerate target HF-like states, as in the case of N$_{2}$O. In this situation, the degenerate HF-like CSFs from the same spin-space symmetry must be combined appropriately to produce a set of orthogonal wave functions. This is achieved most readily, without any changes to the UKRmol+ codes, by contracting all these configurations into a single CI expansion and proceeding as in the case of a CAS calculation.

\section{Conclusions}\label{sec:conclusions}

We have described a set of Perl-based scripts which efficiently drive the complex UKRmol+ suite of codes for electron scattering and photoionization calculations in the fixed-nuclei approximation. The required user input has been condensed, as much as possible, to the physical parameters of the model to reduce the need to understand a number of technical switches and parameters of the calculation. The scripts provide the functionality to automatically generate at run-time the input files for the various programs of the suite and to extract and post-process the most relevant physical outputs of the calculation. 

In addition, the scripts drive the prerequisite quantum chemistry calculations including the open-source codes Psi4 and Molcas. Therefore the full R-matrix calculations described above can be performed using only open-source software. We have provided several basic examples of scattering and photoionization calculations which the users can modify to generate their input for arbitrary molecules.

The scripts have been used for many years to generate the data required to model, for example, nuclear dynamics in resonant electron-molecule scattering~\cite{houfek2016,zawadzki2018,alt2021,rageshkumar2022,dvorak2022a,dvorak2022b}, time-resolved photoelectron spectroscopy~\cite{richter2019,brambila2017} and time-dependent ultrafast laser-molecule interactions~\cite{PRA-RMT-molec,PRA_RABBIT_TI}, to name a few recent applications. The release of these scripts will enable many others in the AMO community to investigate photoionization and ultrafast processes either fully with the UKRmol+ suite or by generating the molecular data (e.g. transition dipoles) required by their own codes and approaches.

\section*{Acknowledgement}
J.~D.~G. is grateful to Daniel Darby-Lewis who initiated the implementation of positron scattering calculations in the scripts supported by eCSE project eCSE13-14. The development of the UKRmol-scripts has been supported by EPSRC under grants EP/P022146/1 and EP/R029342/1. V.~G. acknowledges support from the EPSRC Doctoral Training Partnership EP/T518165/1. Some of the developments were carried out as part of projects 20-15548Y of the Czech Science Foundation and the PRIMUS project (20/SCI/003) of Charles University.

Large-scale deployment of this software has been tested using computational resources supplied by the project “e-Infrastruktura CZ” (e-INFRA LM2018140) provided within the program Projects of Large Research, Development and Innovations Infrastructures. This work was also supported by the Ministry of Education, Youth and Sports of the Czech Republic through the e-INFRA CZ (ID:90140).





\bibliographystyle{elsarticle-num}
\bibliography{rmat}







\end{document}